\documentclass[12pt]{article}
\newcommand{\ba}{\begin{array}}
\newcommand{\ea}{\end{array}}
\newcommand{\nn}{\nonumber\\}

\newcommand{\del}{\partial}

\newcommand{\rar}{\rightarrow}

\newcommand{\fr}{\frac}
\newcommand{\diag}{{\mbox{diag}}}

\newcommand{\scr}{\scriptsize}
\newcommand{\dis}{\displaystyle}

\newcommand{\htil}{\tilde{h}}
\newcommand{\at}{\tilde{a}}
\newcommand{\kt}{\tilde{k}}

\newcommand{\zt}{\tilde{z}}
\newcommand{\wt}{\tilde{w}}
\newcommand{\At}{\tilde{A}}
\newcommand{\Bt}{\tilde{B}}

\newcommand{\phit}{\tilde{\phi}}

\makeatletter
\@addtoreset{equation}{section}

\makeatother
\usepackage{amssymb}

\begin{document}

\begin{titlepage}
\null
\begin{flushright}
%-/-
%\\
hep-th/0601209
\\
January, 2006
\end{flushright}

\vskip 1.5cm
\begin{center}

{\LARGE \bf Noncommutative Ward's Conjecture}

\vskip 0.5cm

{\LARGE \bf and Integrable Systems}

\vskip 1.7cm
\normalsize

{\large Masashi Hamanaka\footnote{The author visits Oxford 
from 16 August, 2005 to  15 August, 2006, supported
by the Yamada Science Foundation.
E-mail: hamanaka@math.nagoya-u.ac.jp, hamanaka@maths.ox.ac.uk}}

\vskip 1.5cm

        {\it Graduate School of Mathematics, Nagoya University,\\
                     Chikusa-ku, Nagoya, 464-8602, JAPAN}
\vskip 0.5cm

        {\it Mathematical Institute, University of Oxford,\\
                     24-29, St Giles', Oxford, OX1 3LB, UK}

\vskip 1.5cm

{\bf \large Abstract}

\end{center}
Noncommutative Ward's conjecture is
a noncommutative version of the original
Ward's conjecture which says that
almost all integrable equations can be obtained
from anti-self-dual Yang-Mills equations by reduction.
In this paper, we prove that
wide class of noncommutative integrable equations
in both $(2+1)$- and  $(1+1)$-dimensions
are actually reductions of noncommutative anti-self-dual Yang-Mills
equations with finite gauge groups,
which include noncommutative versions of
Calogero-Bogoyavlenskii-Schiff eq.,
Zakharov system, Ward's chiral and topological chiral models,
(modified) Korteweg-de Vries,
Non-Linear Schr\"odinger, Boussinesq, N-wave,
(affine) Toda, sine-Gordon, Liouville, Tzitz\'eica,
(Ward's) harmonic map eqs., and so on.
This would guarantee existence of twistor description of them
and the corresponding physical situations in N=2 string theory,
and lead to fruitful applications to noncommutative
integrable systems and string theories. 
Some integrable aspects of them are also discussed.

\end{titlepage}
\clearpage
\baselineskip 6.4mm
\section{Introduction}

Non-commutative (NC) gauge theories are realized as
effective theories of D-branes 
in background of magnetic fields and
have been studied very intensively in string theories.
(For reviews, see \cite{NC}.)
In some contexts, NC solitons are D-branes themselves
and play key roles in the study of D-brane dynamics.
In particular, NC anti-self-dual Yang-Mills (ASDYM) equations
are very important because they are proved to be integrable
in some sense and give rise to $U(1)$ instanton solutions
special to NC theories \cite{NeSc}.
(See also, \cite{YM})

NC extension of soliton theories and integrable systems
is also one of hot topics and naively expected to
be integrable as well and fruitful to yield new physical objects
and to be applied to corresponding physical situations.
(For reviews, see e.g. \cite{Hamanaka_p, Tamassia}.)
However, these equations contain no gauge field 
and the equivalence between extension to NC spaces and
presence of background magnetic fields is non-trivial.
That is why it is very important to
reveal the corresponding physical pictures
of these NC integrable equations.\footnote{
In the present paper,
integrability of NC equations means
existence of infinite number of conserved quantities or
exact multi-soliton solutions though the notion of integrability
has not yet been established on NC ($1+1$)-dimensional space-time.
}

On commutative spaces, there is a famous conjecture
on integrable systems which was proposed
by Richard Ward \cite{Ward, Ward2} and often
called {\it Ward's conjecture}. He conjectured
in \cite{Ward} that

\vspace{3mm}
\noindent
{\it
many (and perhaps all?) of the ordinary and partial
equations that are regarded as being integrable or solvable
may be obtained from the anti-self-dual gauge field equations
(or its generalizations) by reduction.}
\vspace{3mm}

\noindent
This is a very interesting conjecture
which connects various lower-dimensional integrable
equations in scalar theories with a master equation in gauge theories.
Now we know that almost all famous integrable equations
such as Korteweg-de-Vries (KdV) equation
are actually derived from ASDYM equation in $(2+2)$-dimension
by reduction and summarized
in the book of Mason and Woodhouse \cite{MaWo}.
(See also, e.g. \cite{AbCl, AbCh, Reduction}.)
This scenario can be embedded into
geometrical framework of twistor theory and we can understand 
origins of integrability of lower dimensional integrable
equations and classify them in some extent
from the geometrical viewpoints.

ASDYM equations in $(2+2)$-dimension have the corresponding physical pictures
in N=2 string theories \cite{OoVa}
and NC extension of them are successful \cite{LPS}.
Hence if Ward's conjecture still holds on NC spaces,
the reduced NC equations must have physical pictures
and various applications to the corresponding situations
would be possible. Furthermore,
NC twistor theories for NC ASDYM equation have been already
developed by several authors
in \cite{KKO, Takasaki, Hannabuss,Brain, BrHa}\footnote{
The author thanks Brain and Hannabuss for informing him
of their new results \cite{Brain, BrHa}.} for the Euclidean signature,
(See also \cite{HLW}),
and in \cite{IhUh} for the ultrahyperbolic signature,
which would leads to classification of NC integrable equations
as in commutative case.
That is why confirmation of NC Ward's
conjecture is worth studying from the viewpoints of
both N=2 string theories and twistor theories.

NC Ward's conjecture is first proposed in \cite{HaTo}
as a future direction of integrable systems. 
Simple reductions of NC ASDYM equations in relation to
the string theories have been already studied
intensively, and exact soliton solutions are
constructed with application to D-branes dynamics
\cite{HLW, LePo,Bieling,Wolf, LPS2, LePo3, ChLe}.
This suggests that all other NC integrable equations would
also have physical situations and might lead to various successful
applications to the corresponding D-brane
dynamics and so on. Furthermore,
several nontrivial reductions of NC ASDYM equations
have been successful for NC NLS eq. \cite{Legare}, NC supersymmetric
KdV eq. \cite{Legare, NiRa}, NC Burgers eq. \cite{HaTo2},
NC sine-Gordon eq. \cite{GMPT, LMPPT},
NC (affine) Toda and NC Liouville equations \cite{Cabrera-Carnero},
NC KdV, N-wave, Kadomtsev-Petviashvili (KP), Davey-Stewartson (DS)
equations \cite{Hamanaka2}.
All of the reduced equations have
integrable-like properties, such as,
linealizability, existence of infinite conserved quantities
or exact multi-soliton solutions and so on,
which might be explained clearly
in the geometric framework of NC twistor theories.

These results on NC Ward's conjecture have been
examined individually by focusing on each equation and
arguments with  {\it symmetry} in the reduction are missing.
Furthermore, most of the reduced equations belong to
($1+1$)-dimension,\footnote{
In \cite{Hamanaka2}, several NC equations in ($2+1$)-dimension,
such as KP and DS equations, are derived from NC ASDYM equation
with {\it infinite} gauge groups. These reductions have no naive
twistor description on commutative spaces and would not be
considered to be examples of Ward's conjecture.}
in which integrability is hard to discuss
and even to define because noncommutativity must be
introduced into time direction.

In this paper, we show that
NC integrable equations in both $(2+1)$- and  $(1+1)$-dimensions
are actually reductions of noncommutative anti-self-dual Yang-Mills
equations with {\it finite} gauge groups,
which include new non-trivial reductions
to Calogero-Bogoyavlenskii-Schiff (CBS) eq.
and Zakharov system in $(2+1)$-dimension
and mKdV, Boussinesq, Tzitz\'eica and harmonic map eqs.
and topological chiral model in $(1+1)$-dimension.
The discussion is clarified from the viewpoint of
symmetry reduction in some extent.
This would lead to a systematic classification of NC
integrable systems in the framework of NC
twistor theory and to fruitful applications to
the corresponding situations in N=2 string theory.
Now we can say that NC Ward's conjecture
has been almost confirmed by the explicit examples.
We also show gauge equivalence of reductions
to potential KdV, KdV and modified KdV (mKdV) eqs.
and develop NC B\"acklund transformation for NC ASDYM
equation in some extent, which would be applied to
lower-dimensional equations by reduction.

This paper is organized as follows.
In section 2, we discuss some aspects of 
of NC ASDYM equations including Yang's forms
and B\"acklund transformations and summarize
the strategy of reductions of the NC ASDYM equations.
In section 3, we perform reductions of NC ASDYM
to NC integrable equations $(2+1)$-dimension.
In section 4, we give various examples of
NC Ward's conjecture into $(1+1)$-dimension.
We comment on the integrable
properties of the reduced equations at each example of the reduction.
In section 5, we conclude and discuss further directions.

\section{NC ASDYM equation}

In this section, we review some aspects of
NC ASDYM equation and establish notations.
Discussion on NC B\"acklund transformation is new.

\subsection{NC gauge theory}

NC spaces are defined
by the noncommutativity of the coordinates:
\begin{eqnarray}
\label{nc_coord}
[x^i,x^j]=i\theta^{ij},
\end{eqnarray}
where $\theta^{ij}$ are real constants
called the {\it NC parameters}. 
The NC parameter is anti-symmetric with respect to $i,j$:
$\theta^{ji}=-\theta^{ij}$ and the rank is even.
This relation looks like the canonical commutation
relation in quantum mechanics
and leads to ``space-space uncertainty relation.''
Hence the singularity which exists on commutative spaces
could resolve on NC spaces.
This is one of the prominent features of NC
field theories and yields various new physical objects.

NC field theories are given by the exchange of ordinary products
in the commutative field theories for the star-products and
realized as deformed theories from the commutative ones. 
The ordering of non-linear terms are determined by
some additional requirements such as gauge symmetry.
The star-product is defined for ordinary fields on commutative
spaces. For Euclidean spaces, it is explicitly given by
\begin{eqnarray}
f\star g(x)&:=&{\mbox{exp}}
\left(\frac{i}{2}\theta^{ij} \partial^{(x^{\prime})}_i
\partial^{(x^{\prime\prime})}_j \right)
f(x^\prime)g(x^{\prime\prime})\Big{\vert}_{x^{\prime}
=x^{\prime\prime}=x}\nonumber\\
&=&f(x)g(x)+\frac{i}{2}\theta^{ij}\partial_i f(x)\partial_j g(x)
+O (\theta^2),
\label{star}
\end{eqnarray}
where $\del_i^{(x^\prime)}:=\del/\del x^{\prime i}$
and so on.
This explicit representation is known
as the {\it Moyal product} \cite{Moyal}.
The star-product has associativity:
$f\star(g\star h)=(f\star g)\star h$,
and returns back to the ordinary product
in the commutative limit:  $\theta^{ij}\rar 0$.
The modification of the product  makes the ordinary
spatial coordinate ``noncommutative,''
that is, $[x^i,x^j]_\star:=x^i\star x^j-x^j\star x^i=i\theta^{ij}$.

We note that the fields themselves take c-number values
as usual and the differentiation and the integration for them
are well-defined as usual, for example, $\del_i\star \del_j
=\del_i\del_j,~$ and the wedge product of $\omega=\omega_i(x)dx^i$
and $\lambda=\lambda_j(x)dx^j$ is $\omega_i\star \lambda_j dx^i\wedge dx^j$.

NC gauge theories are defined in this way
by imposing NC version of gauge symmetry,
where the gauge transformation is defined as follows:
\begin{eqnarray}
 A_\mu \rightarrow g^{-1}\star A_\mu\star g +g^{-1}\star \del_\mu g,
\end{eqnarray}
where $g$ is an element of the gauge group $G$.
This is sometimes called the {\it star gauge transformation} \cite{NC}.
We note that because of noncommutativity,
the commutator terms in field strength are always needed
even when the gauge group is abelian in order to preserve
the star gauge symmetry. This $U(1)$ part of the gauge group
actually plays crucial roles in general.
We note that because of noncommutativity of matrix elements,
cyclic symmetry of traces is broken in general:
\begin{eqnarray}
 \label{trace}
 {\mbox{Tr }}A\star B\neq  {\mbox{Tr }}B\star A.
\end{eqnarray}
Therefore, gauge invariant quantities becomes hard to
define on NC spaces.

\subsection{NC ASDYM equation}

Let us consider Yang-Mills theories on 
$(2+2)$-dimensional NC spaces whose 
real coordinates of the space are denoted by $(x^0,x^1,x^2,x^3)$,
where the gauge group is $GL(N,\mathbb{C})$.
Here, we follow the convention in \cite{MaWo} as follows.

First, we introduce double null coordinates
of $(2+2)$-dimensional space as follows
\begin{eqnarray}
ds^2=2(dzd\tilde{z}-dwd\tilde{w}),
\end{eqnarray}
where
       \begin{eqnarray}
        \left(\begin{array}{cc}\tilde{z}&w\\\tilde{w}&z\end{array}\right)
        =\frac{1}{\sqrt{2}}
         \left(\begin{array}{cc}x^0+ix^1&x^2-ix^3\\x^2+ix^3&x^0-ix^1
               \end{array}\right).
       \end{eqnarray}
The reality conditions are $\bar{z}=\tilde{z},
\bar{w}=\tilde{w}$.\footnote{
For Euclidean signature, we have only to put another
reality condition as $\bar{z}=\tilde{z},
\bar{w}=-\tilde{w}$.}
The coordinate vectors $\del_z,\del_z.\del_{\wt}, \del_{\zt}$
 form a null tetrad and are represented explicitly as:
 \begin{eqnarray}
  \label{tetrad}
&&  \del_z=\frac{1}{\sqrt{2}}\left(\frac{\del}{\del x^0}+i\frac{\del}{\del
                            x^1}\right),~~~
  \del_{\tilde{z}}=\frac{1}{\sqrt{2}}
  \left(\frac{\del}{\del x^0}-i\frac{\del}{\del
                            x^1}\right),\nn
&&  \del_w=\frac{1}{\sqrt{2}}\left(\frac{\del}{\del x^2}+i\frac{\del}{\del
                            x^3}\right),~~~
  \del_{\tilde{w}}=\frac{1}{\sqrt{2}}
  \left(\frac{\del}{\del x^2}-i\frac{\del}{\del
                            x^3}\right).
 \end{eqnarray}
For ultrahyperbolic signature, Hodge dual operator $*$
satisfies $*^2=1$ and hence the space of 2-forms $\beta$
decomposes into the direct sum of eigenvalues of $*$ with eigenvalues $\pm 1$,
that is, self-dual (SD) part ($*\beta=\beta$)
and anti-self-dual (ASD) part ($*\beta=-\beta$). 

Typical examples of SD forms are
\begin{eqnarray}
 \alpha=dw\wedge d{z},~~~\tilde{\alpha}=d\wt\wedge d\zt,
  ~~~\omega=dw\wedge d\wt-dz\wedge d\zt,
  \label{omega}
\end{eqnarray}
and those of ASD forms are
\begin{eqnarray}
 dw\wedge d{\zt},~~~d\wt\wedge dz,~~~dw\wedge d\wt+dz\wedge d\zt.
\end{eqnarray}

NC ASDYM equation is derived from compatibility condition
of the following linear system:
\begin{eqnarray}
L\star \Psi&:=&(D_w-\lambda D_{\tilde{z}})\star \Psi=
 \left(\del_w+A_w-\lambda (\del_{\zt}+A_{\tilde{z}})\right)\star \Psi(x;\lambda)
= 0,\nn
M\star \Psi&:=&(D_z-\lambda D_{\tilde{w}})\star \Psi=
  \left(\del_z+A_z-\lambda (\del_{\wt}+A_{\tilde{w}})\right)\star \Psi(x;\lambda)
= 0,
\label{lin_asdym}
\end{eqnarray}
where $A_z,A_w,A_{\tilde{z}},A_{\tilde{w}}$
and $D_z,D_w,D_{\tilde{z}},D_{\tilde{w}}$
denote gauge fields and covariant derivatives in the Yang-Mills theory,
respectively. The constant $\lambda$ is called
the {\it spectral parameter}. 
The compatible condition $[L,M]_\star=0$ gives rise to
a quadratic polynomial of $\lambda$ and each coefficient
yields NC ASDYM equations whose explicit representations are as follows:
\begin{eqnarray}
 F_{wz}&=&\del_{w} A_z -\del_{z} A_w+[A_w,A_z]_\star =0,\nn
 F_{\wt\zt}&=&\del_{\tilde{w}} A_{\tilde{z}} -\del_{\tilde{z}} A_{\tilde{w}}
 +[A_{\tilde{w}},A_{\tilde{z}}]_\star =0,\nn
 F_{z\zt}-F_{w\wt}&=&\del_{z} A_{\tilde{z}} -\del_{\tilde{z}} A_{z}
 +\del_{\tilde{w}} A_{w} -\del_{w} A_{\tilde{w}}
 +[A_z,A_{\tilde{z}}]_\star
 -[A_{w},A_{\tilde{w}}]_\star=0.
\label{asdym}
\end{eqnarray}

As in commutative case, if $\Pi$ is a null 2-plane in space-time,
a tangent bivector $\pi$ associated to $\Pi$ satisfies
$\pi_{\mu\nu}\pi^{\mu\nu}=0$ and $\pi_{\mu\nu}dx^\mu\wedge dx^\nu$
is either SD or ASD. 
For example, vector fields $l=\del_w-\lambda \del_{\zt}$
and $m=\del_z-\lambda\del_{\wt}$ form a SD bivector $l\wedge m$.
A 2-plane $\Pi$ is called an {\it $\alpha$-plane} when
the associated tangent bivector $\pi$ is SD.
NC ASDYM eqs are equivalent to the condition that
the connection is flat on the $\alpha$-plane: $F(l,m)=0$.
(For detailed discussion, see e.g. \cite{MaWo}.)

\subsection{NC Yang's equation and $J, K$-matrices}

Here we discuss the potential forms of NC ASDYM equations
such as NC $J,K$-matrix formalisms and NC Yang's equation,
which is already presented by e.g. K.~Takasaki \cite{Takasaki}.
(See also \cite{Yangsform}.)

Let us first discuss the {\it $J$-matrix formalism}
of NC ASDYM equation.
The first equation of NC ASDYM equation
(\ref{asdym}) is the compatible condition
of $D_z\star h=0, D_w\star h=0$,
where $h$ is a $N\times N$ matrix.
Hence we get
\begin{eqnarray}
A_{z}=-h_{z}\star h^{-1}, ~~~  A_{w}=-h_{w}\star h^{-1},
\end{eqnarray}
where $h_z:=\del h/\del z,~h_w:=\del h/\del w$.
Similarly, the second eq. of NC ASDYM equation (\ref{asdym}) leads to
\begin{eqnarray}
A_{\zt}=-\htil_{\zt}\star \htil^{-1}, ~~~
A_{\wt}=-\htil_{\wt}\star \htil^{-1},
\end{eqnarray}
where $\htil$ is also a $N\times N$ matrix.
By defining $J=\htil^{-1}\star h $, 
the third eq. of NC ASDYM equation (\ref{asdym}) becomes NC Yang's equation 
\begin{eqnarray}
\label{yang}
 \del_z(J^{-1} \star \del_{\zt} J)-\del_w (J^{-1}\star \del_{\wt} J)=0.
\end{eqnarray}
%or equivalently,
%\begin{eqnarray}
% \del\left(J^{-1}\star \delt J\right)\wedge \omega=0,
%\end{eqnarray}
%where $\del=dw \del_w +dz\del_z,~\delt=d\wt \del_{\wt} +d\zt\del_{\zt}$
%and $\omega$ is the same one in Eq. (\ref{omega}). 
This representation is useful for
reductions to NC Ward's chiral models and 
NC sine-Gordon, Liouville and Tzitz\'eica equations, and so on.

There is another potential form of NC ASDYM equation,
known as the {\it $K$-matrix formalism}. Under the gauge $A_w=A_z=0$,
the third eq. of NC ASDYM equation (\ref{asdym})
becomes $\del_z A_{\zt}-\del_w A_{\wt}=0$,
which implies existence of a potential $K$ such
that $A_{\zt}=\del_{w}K,A_{\wt}=\del_{z}K$.
Then the second eq. of NC ASDYM equation (\ref{asdym}) becomes
\begin{eqnarray}
 \del_z\del_{\zt}K -\del_w\del_{\wt}K +[\del_w K, \del_z K]_\star=0.
\end{eqnarray}
%The corresponding action is
%\begin{eqnarray}
% S_K=\int
%  {\mbox{Tr} } \left(\del_\mu K \star \del^\mu K
%       +\frac{4}{3}K\star [\del_z K, \del_w K]_\star \right)
% d\wt d\zt dw dz.
%\end{eqnarray}

\subsection{NC B\"acklund transformation}

Here we make a brief discussion on
hidden symmetry behind Yang's equation and
NC B\"acklund transformation for NC ASDYM equation.
This would relates to NC B\"acklund transformation 
for NC integrable equations in lower dimensions via reductions.

First let us extend the discussion on
hidden symmetry for Yang's equation \cite{MaWo}
to NC spaces. A generic $N\times N$ matrix $J$ can
be decomposed as follows:
\begin{eqnarray}
 J=\left(\begin{array}{cc}A^{-1}-\Bt\star \At \star B&-\Bt\star \At
   \\ \At \star B &\At\end{array}
   \right)
 =\left(\begin{array}{cc}1&\Bt
   \\ 0 &\At^{-1}\end{array}
   \right)^{-1}\star
 \left(\begin{array}{cc}A^{-1}&0
   \\ B&1\end{array}
   \right), 
\end{eqnarray}
where $A,\At,B$ and $\Bt$ are $k\times k, \kt\times \kt, \kt \times k$
and $k\times \kt$ matrices, respectively, with $k+\kt=N$
and $A$ and $\At$ are non-singular.
With this decomposition, NC Yang's equation (\ref{yang}) becomes
\begin{eqnarray}
  \del_{z}(A\star \Bt_{\zt} \star \At)-\del_{w}(A\star \Bt_{\wt}\star
  \At)=0,~~~
  \del_{\zt}(\At\star B_z \star A)-\del_{\wt}(\At\star B_w\star A)=0,\nn
  \del_{z}(\At^{-1}\star\At_{\zt})\star \At^{-1}-
  \del_w(\At^{-1}\star \At_{\wt})\star \At^{-1}+
  B_z\star A\star \Bt_{\zt}-B_w\star A\star \Bt_{\wt}=0,\nn
  A^{-1}\star \del_{z}(A_{\zt}\star A^{-1})-
  A^{-1}\star \del_{w}(A_{\wt}\star \At^{-1})+
  \Bt_{\zt}\star \At\star B_{z}-\Bt_{\wt}\star \At\star B_{w}=0.
\label{dYang}
\end{eqnarray}
The first two equations can be interpreted as integrability
conditions. Hence there exist  $k\times \kt$ and $\kt \times k$
matrices $B^{\mbox{\scr{new}}}$ and $\Bt^{\mbox{\scr{new}}}$,
respectively, such that
\begin{eqnarray}
 \label{new}
 \del_{z} B^{\mbox{\scr{new}}}=A\star \Bt_{\wt}\star \At,~~~
  \del_{w} B^{\mbox{\scr{new}}}=A\star \Bt_{\zt}\star \At,\nn  
\del_{\zt} \Bt^{\mbox{\scr{new}}}=\At\star B_w\star A,~~~
  \del_{\wt}\Bt^{\mbox{\scr{new}}}=\At\star B_z\star A.
\end{eqnarray}
Furthermore, if we put
\begin{eqnarray}
\label{new2}
&&A^{\mbox{\scr{new}}}=\At^{-1},~\At^{\mbox{\scr{new}}}=A^{-1},\nn
&&k^{\mbox{\scr{new}}}=\kt,~\kt^{\mbox{\scr{new}}}=k,
\end{eqnarray}
then we obtain a new solution $J^{\mbox{\scr{new}}}$ of NC Yang's equation.
This can be considered as a transformation
$i_k: J\rightarrow J^{\mbox{\scr{new}}}$.
We can see that $i_{n-k}\circ i_k$ is identity.

This can be understood from the viewpoint of
NC B\"acklund transformation for NC ASDYM equation
in the following way. If we take
\begin{eqnarray}
 h=
 \left(\begin{array}{cc}A^{-1}&0
   \\ B&1\end{array}
   \right),~~~
 \htil=\left(\begin{array}{cc}1&\Bt
   \\ 0 &\At^{-1}\end{array}
   \right),
\end{eqnarray}
then we can choose the gauge so that
\begin{eqnarray}
A_{z}=-h_{z}\star h^{-1}, ~~~  A_{w}=-h_{w}\star h^{-1},~~~
A_{\zt}=-\htil_{\zt}\star \htil^{-1}, ~~~
A_{\wt}=-\htil_{\wt}\star \htil^{-1},
\end{eqnarray}
and the linear systems of NC ASDYM is
\begin{eqnarray}
 \label{orig}
 L&=&\del_w-\lambda \del_{\zt}+
   \left(\begin{array}{cc}A^{-1}\star A_w&\lambda\Bt_{\zt}\star \At
   \\ -B_w\star A&-\lambda\At^{-1}\star \At_{\zt}\end{array}
   \right),\nn
 M&=&\del_z-\lambda \del_{\wt}+
   \left(\begin{array}{cc}A^{-1}\star A_z&\lambda\Bt_{\wt}\star \At
   \\ -B_z\star A&-\lambda\At^{-1}\star \At_{\wt}\end{array}
   \right).
\end{eqnarray}
Now let us consider the following gauge transformation
\begin{eqnarray}
 \label{irr}
 L^{\mbox{\scr{new}}}=g^{-1}\star L\star g,~~~
 M^{\mbox{\scr{new}}}=g^{-1}\star M\star g,~~~  
\end{eqnarray}
where
\begin{eqnarray}
 g^{-1}= \left(\begin{array}{cc}0&\lambda\At
   \\ -A&0\end{array}
   \right).
\end{eqnarray}
The transformed linear systems are as follows
\begin{eqnarray}
 L^{\mbox{\scr{new}}}&=&\del_w-\lambda \del_{\zt}+
   \left(\begin{array}{cc}-\At_{\zt}\star \At^{-1}&\lambda\At \star B_{w}
   \\ -A\star \Bt_{\zt}&\lambda A_{\zt}\star A^{-1}\end{array}
   \right),\nn
 M^{\mbox{\scr{new}}}&=&\del_z-\lambda \del_{\wt}+
   \left(\begin{array}{cc}-\At_{\wt}\star \At^{-1}&\lambda\At \star B_z
   \\ -A\star \Bt_{\wt}&\lambda A_{\wt}\star A^{-1}\end{array}
   \right).
\end{eqnarray}
Now by making the identification just as (\ref{new}) and (\ref{new2}),
$L^{\mbox{\scr{new}}}$ and $M^{\mbox{\scr{new}}}$ 
have the same form as Eq. (\ref{orig}).
Hence the hidden symmetry can be interpreted as
a B\"acklund transformation for the linear systems
of NC ASDYM equation.

We note that the gauge transformation (\ref{irr})
is irregular in the sense that
the determinant of $g$ becomes singular at $\lambda=0, \infty$.
However, we can treat this point properly as in commutative case
because the spectral parameter $\lambda$ is a commutative variable.
(See also \cite{Takasaki}.)
More detailed discussion will be reported in a separated paper.
(For commutative discussion, see e.g. \cite{CMN, MaWo}.
B\"acklund transformation for a few NC integrable equations
is briefly discussed in e.g. \cite{DiMH}.)

\subsection{Reduction of NC ASDYM equation}

Here we summarize the strategy for
reductions of NC ASDYM equation into lower-dimensions.
Reductions are classified by the following ingredients:
\begin{itemize}
 \item A choice of gauge group
 \item A choice of symmetry, such as, translational symmetry
 \item A choice of gauge fixing
 \item A choice of constants of integrations in the process of
       reductions
\end{itemize}
Gauge groups are in general $GL(N,\mathbb{C})$.
We have to take $U(1)$ part into account in NC case.
A choice of symmetry reduces
NC ASDYM equations to simple forms.
We note that noncommutativity must be eliminated
in the reduced directions
because of compatibility with the symmetry.
Hence within the reduced directions, discussion about the
symmetry is the same as commutative one.
A choice of gauge fixing
is the most important ingredient in this paper
which is shown explicitly at each subsection.
The residual gauge symmetry sometimes shows equivalence of
a few reductions as we will see in Secs. 4.1 and 4.2.
Constants of integrations in the process of reductions
sometimes lead to parameter families of NC reduced equations,
however, in this paper, we set all integral constants
zero for simplicity. 

In the following sections, we present explicit reductions
of NC ASDYM equation to various NC integrable equation classified
in the above viewpoints. 

\vspace{3mm}

Before discussing the reductions,
we comment on how to guess the reduction of NC ASDYM equation
to the relevant lower-dimensional NC integrable equations.
Because the original ASDYM equation
possesses the Lax representation in a wider sense,
the reduced equations always possess
the Lax representations.
The Lax representations are common in many integrable
equations. That is one of the reasons
why Ward's conjecture is reasonable.
Oppositely, in some case, Lax representations of
integrable systems with a
spectral parameter in $(1+1)$-dimensions
can be embedded into Lax form of ASDYM equation.
In order to see this, for example,
let us consider a Lax formalism
of NC KdV equation:
\begin{eqnarray}
P\star \psi=
(\del_x^2+u)\star \psi=
\lambda \psi,
 ~~~B\star \psi=
(\del_t-\del_x^3-\frac{3}{2}u\del_x-\frac{3}{4}u')\star \psi
=0,
 \label{Lax}
\end{eqnarray}
where $u':=\del u/\del x, \dot{u}:=\del u/\del t$.
The compatibility condition $[P,B]_\star=0$ gives rise
to NC KdV equation $4\dot{u}-u^{\prime\prime\prime}
+3u'\star u+3u\star u'=0$.

Now let us introduce
\begin{eqnarray}
 \Psi=\left(\begin{array}{c}\psi\\\psi_x\end{array}\right),
\end{eqnarray}
then the derivatives of $\Psi$ with respect to $x$ and $t$
are calculated from Eq. (\ref{Lax}) as follows 
\begin{eqnarray}
 \del_x \Psi&=&\left\{
 \left(\begin{array}{cc}0&1\\-u&0\end{array}\right)
 +\lambda\left(\begin{array}{cc}0&0\\1&0\end{array}\right)
 \right\}
\left(\begin{array}{c}\psi\\\psi_x\end{array}\right),\\
 \label{s}
  \del_t\Psi&=&
   \left\{\frac{1}{4}
 \left(\begin{array}{cc}-u'&2u\\-u^{\prime\prime}-2u\star u&u'\end{array}\right)
 +\lambda\left(\begin{array}{cc}0&1\\-(1/2)u&0\end{array}\right)
 +\lambda^2\left(\begin{array}{cc}0&0\\1&0\end{array}\right)
 \right\}
\left(\begin{array}{c}\psi\\\psi_x\end{array}\right),\nn
&=&
   \left\{\frac{1}{4}
 \left(\begin{array}{cc}-u'&2u\\-u^{\prime\prime}-2u\star u&u'\end{array}\right)
 +\lambda\left[\del_x+
                \left(\begin{array}{cc}0&0\\(1/2)u&0\end{array}\right)\right]
 \right\}
\left(\begin{array}{c}\psi\\\psi_x\end{array}\right).
\label{t}
\end{eqnarray}
We note that it is nontrivial that
Eq. (\ref{t}) becomes linear with respect to $\lambda$
as in the second line of Eq. (\ref{t}).
{}From this representation,
we can find a candidate for a reduction to NC KdV equation of 
NC ASDYM equation by identifying the linear systems (\ref{lin_asdym}) with
Eqs. (\ref{s}) and (\ref{t}): 
\begin{eqnarray*}
A_{\tilde{w}}=\left(\begin{array}{cc}0&0\\u/2&0\end{array}\right), 
A_{\tilde{z}}=\left(\begin{array}{cc}0&0\\1&0\end{array}\right),
A_w=\left(\begin{array}{cc}0&-1\\u&0\end{array}\right),
A_z=\frac{1}{4}
\left(\begin{array}{cc} u^\prime& -2u \\
 u^{\prime\prime}+2u\star u &-u^\prime
\end{array}\right),
\end{eqnarray*}
together with $\del_w=\del_{\wt}=\del_x$ and $\del_{\zt}=0$.
This actually coincides with the reduction condition
to NC KdV equation from NC ASDYM equation as in Eq. (\ref{kdv2}).

\section{Reduction to $(2+1)$-dimension}

In this section, we present non-trivial
reductions of NC ASDYM equation to $(2+1)$-dimension.
Reductions to NC CBS equation and NC Zakharov system are
new results. 

First, let us take a reduction by a null translation:
\begin{eqnarray}
 Y=\del_{\tilde{z}}.
\end{eqnarray}
The gauge field $A_{\tilde{z}}$ becomes
a Higgs field which is denoted by $\Phi_{\tilde{z}}$.

For $G=GL(2,\mathbb{C})$,
standard choices of  gauge fixing of $\Phi_{\tilde{z}}$ are as follows:
\begin{eqnarray*}
\mbox{(i)} ~~~\Phi_{\tilde{z}}=\left(
                        \begin{array}{cc}0&0\\1&0
                        \end{array}\right)~~~~~~
\mbox{(ii)}~~~\Phi_{\tilde{z}}=\kappa \left(
                        \begin{array}{cc}1&0\\0&-1
                        \end{array}\right),
\end{eqnarray*}
where $\kappa$ is a constant.
By taking a gauge $A_{\tilde{w}}=0$, we get a reduced
NC ASDYM equation:
\begin{eqnarray}
\label{asdym2}
&&\del_{\tilde{w}} A_{w}
 +[A_z,\Phi_{\tilde{z}}]_\star=0,\\
&&\del_{w} A_z -\del_{z} A_w+[A_w,A_z]_\star =0.
\label{asdym3}
\end{eqnarray}

In the following, we will see the cases (i) and (ii)
give rise to NC CBS equation and  NC Zakharov system
which are $(2+1)$-dimensional generalizations of
NC KdV equation and NC NLS equation, respectively.

\subsection{Reduction to NC CBS equation}

The Calogero-Bogoyavlenskii-Schiff (CBS) equation\footnote{
As far as the author knows,
this equation was obtained first by Calogero \cite{Calogero},
and Bogoyavlenskii \cite{Bogoyavlenskii},
and Schiff \cite{Schiff} from different viewpoints,
probably independently. That is why we use the present name.}
is a $(2+1)$-dimensional generalization of the KdV equation
and one of the important equations in $(2+1)$-dimension.
NC version of it was first derived in \cite{Toda}.
Here, we derive the equation from NC ASDYM equation
by reduction as follows. For the commutative discussion,
see \cite{MaWo,Schiff}.

Now let us take further reduction
on the gauge fields in the ASDYM
equation (\ref{asdym2}) and (\ref{asdym3}) as follows:
\begin{eqnarray*}
A_{w}=\left(\begin{array}{cc}q&~-1\\q_w+q\star q
&~-q\end{array}\right),~
A_{z}=\left(\begin{array}{cc}
\displaystyle (1/2)q_{w\tilde{w}}+
q_{\tilde{w}}\star q+\alpha
&-q_{\tilde{w}}\\
\phi&
\displaystyle -(1/2)q_{w\tilde{w}}
-q\star q_{\tilde{w}}+\alpha
            \end{array}\right),
\end{eqnarray*}
where $\alpha$ and $\phi$ are differential polynomials of $q$.
The gauge group is $GL(2,\mathbb{R})$.
The ordering of nonlinear terms in the trace part of $A_z$
is determined in order that in the case of $\del_w=\del_{\wt}$,
this reduction to NC CBS equation
should coincide with that to NC KdV equation.
The unknown variable $\alpha$ also satisfies the condition that 
$\alpha$ should vanish in the both cases of the commutative
limit and $\del_w=\del_{\tilde{w}}$.

%This conditions reduces to those
%in Mason-Sparling's paper \cite{MaSp}
%in the commutative limit.

Eq. (\ref{asdym2}) is trivially satisfied.
Eq. (\ref{asdym3}) yields
\begin{eqnarray}
\label{1bcs}
&&\phi=\displaystyle
-q_z+\frac{1}{2}q_{ww\tilde{w}}
+\frac{1}{2}\left\{q, q_{w\tilde{w}}\right\}_\star
+q\star q_{\tilde{w}}\star q
+2q_{\tilde{w}}\star q_w+\alpha_w+[q,\alpha]_\star,
\\
\label{2bcs}
&&\phi=\displaystyle
-q_z+\frac{1}{2}q_{ww\tilde{w}}
+\frac{1}{2}\left\{q, q_{w\tilde{w}}\right\}_\star
+q\star q_{\tilde{w}}\star q
+2q_w\star q_{\tilde{w}}-\alpha_w+[q,\alpha]_\star,
\\
\label{3bcs}
&&(\phi-q_z)_w -\left\{q, \phi+ q_z\right\}_\star
+[q_w+q\star q,\alpha]_\star
+q_w\star q_{\tilde{w}}\star q
+q\star  q_{\tilde{w}}\star q_w
\nn
&&~~~+\displaystyle\frac{1}{2}\left\{q_w, q_{w\tilde{w}}\right\}_\star
+\displaystyle\frac{1}{2}\left\{q\star q,  q_{w\tilde{w}}\right\}_\star
+\left\{q, q\star q_{\tilde{w}}\star q\right\}_\star
=0. 
\label{red_kdv3}
\end{eqnarray}
where $\left\{A,B\right\}_\star :=A\star B +B\star A$.

{}From Eqs. (\ref{1bcs}) and (\ref{2bcs}),
we get $\alpha=\del_w^{-1}[q_w, q_{\tilde{w}}]_\star$,
where $\del_w^{-1} f(w):=\int^w dw^\prime f(w^\prime)$.
This satisfies the condition for $\alpha$.

Then we get
\begin{eqnarray}
\phi=\displaystyle
-q_z+\frac{1}{2}q_{ww\tilde{w}}
+\frac{1}{2}\left\{q,q_{w\tilde{w}}\right\}_\star
+\frac{1}{2} \left\{q_{w}, q_{\tilde{w}}\right\}_\star
+q\star q_{\tilde{w}}\star q
+[q,\del_w^{-1}[q_w, q_{\tilde{w}}]_\star]_\star,
\end{eqnarray}
and Eq. (\ref{3bcs}) becomes NC potential CBS equation:
\begin{eqnarray}
 q_{zw}=\displaystyle
  \frac{1}{4}q_{www\tilde{w}}
+\left\{q_w, q_{w\tilde{w}}\right\}_\star
+\frac{1}{2} \left\{q_{\tilde{w}}, q_{ww}\right\}_\star
+[q_w,\del_w^{-1}[q_w, q_{\tilde{w}}]_\star]_\star,
\label{pbcs}
\end{eqnarray}
which is derived from the NC CBS equation \cite{Toda, HaTo}
\begin{eqnarray}
 u_z=\displaystyle\frac{1}{4}u_{ww\tilde{w}}
+\frac{1}{2}\left\{u, u_{\tilde{w}}\right\}_\star
+\frac{1}{4} \left\{u_{\tilde{w}}, \del_w^{-1} u_{\tilde{w}}
             \right\}_\star
+\frac{1}{4}\del_w^{-1}[u,
\del_w^{-1}[u, \del_w^{-1} u_{\tilde{w}}]_\star]_\star
\label{bcs} 
\end{eqnarray}
by setting $2q_w =u$.

If we impose an additional symmetry described by non-null translation
$X=\del_w-\del_{\tilde{w}}$, which is realized by the
identification $\del_w=\del_{\tilde{w}}$,
the present discussion is reduced to that to NC KdV equation
\cite{Hamanaka2}, which is NC generalization of Mason-Sparling's result
\cite{MaSp}. In this sense, NC CBS equation is a $(2+1)$-dimensional
extension of NC KdV equation \cite{Toda, HaTo}.

\subsection{Reduction to NC Zakharov system}

Zakharov system \cite{Zakharov} is a $(2+1)$-dimensional
extension of NLS equation in the same sense as CBS equation.
Here we take the choice (ii) for $\Phi_{\zt}$,
and make the same procedure to yield
NC Zakharov system from NC ASDYM equation as NC CBS equation.
For the commutative discussion,
see \cite{MaWo,Strachan0}.

Here let us take another further reduction
on the gauge fields in the ASDYM equation
(\ref{asdym2}) and 
(\ref{asdym3}) as follows:
\begin{eqnarray}
A_{w}=\left(\begin{array}{cc}0&\psi\\\tilde{\psi}&0\end{array}\right),~
A_{z}=\frac{1}{2\kappa}\left(\begin{array}{cc}
U&\psi_{\tilde{w}}\\
-\tilde{\psi}_{\tilde{w}}&
V\end{array}\right),
\end{eqnarray}
where $U$ and $V$ are differential polynomials
of $\psi$ and $\tilde{\psi}$. The off-diagonal
elements in $A_z$ are determined in order that
Eq. (\ref{asdym2}) should be solved.
Each component of Eq. (\ref{asdym3}) becomes
\begin{eqnarray*}
&&U_w-\psi \star \tilde{\psi}_{\tilde{w}}
-\psi_{\tilde{w}}\star
\tilde{\psi}=0,~V_w+\tilde{\psi}\star \psi_{\tilde{w}}
+\tilde{\psi}_{\tilde{w}}\star \psi=0,\\
&&
\psi_{w\tilde{w}}-2\kappa \psi_z
+\psi \star V-U\star \psi=0,~
-\tilde{\psi}_{w\tilde{w}}-2\kappa \psi_z
+\tilde{\psi} \star U-V\star \tilde{\psi}=0.
\end{eqnarray*}
The first two equations lead to
$U=\del_w^{-1}\del_{\tilde{w}}(\psi \star \tilde{\psi}),~
V=-\del_w^{-1}\del_{\tilde{w}}(\tilde{\psi}\star \psi)$.
By taking $\dis \kappa=i/2, 
\tilde{\psi}=\varepsilon\bar{\psi}$ where $\varepsilon=\pm 1$,
the remaining two equations
coincide with NC Zakharov system: 
\begin{eqnarray}
 i\psi_z&=&\psi_{w\tilde{w}}-\varepsilon \psi \star
  \del_w^{-1}\del_{\bar{w}}(\bar{\psi} \star \psi)
  -\varepsilon \del_w^{-1}\del_{\bar{w}}(\psi\star \bar{\psi})\star \psi,
% i\bar{\psi}_z&=&-\bar{\psi}_{w\bar{w}}
%  \pm\bar{\psi} \star
%  \del_w^{-1}\del_{\bar{w}}(\psi \star \bar{\psi})
%  \pm\del_w^{-1}\del_{\bar{w}}(\bar{\psi}\star \psi)\star \bar{\psi}.
\end{eqnarray}
and the complex conjugate.
Now the gauge group reduces to $U(1,1)$ or $U(2)$.
The present discussion reduces to that to NC NLS equation
by identifying $\del_w=\del_{\wt}$.
This system is studied in more general framework
by Dimakis and M\"uller-Hoissen
and proved to possess infinite
conserved quantities \cite{DiMH4}
in terms of Strachan's product \cite{Strachan}.

Zakharov system can be embedded into framework of a generalized twistor
theory \cite{Strachan2}. The NC extension is worth studying.

\subsection{Reduction to NC Ward's chiral model}

Here let us start with NC Yang's Eq. (\ref{yang})
and take a simple dimensional reduction by
non-null transformation such as $X=\del_w-\del_{\wt}$.
The reduced equation is NC Ward's chiral model:
\begin{eqnarray}
 (\eta^{ij}+\epsilon^{ij2})\del_i(J^{-1}\star \del_j J)=0,
\end{eqnarray}
where $\eta^{ij}$ ($i,j=0,1,2$) is the Minkowski metric:
$\eta^{ij}=\diag (+,+,-)$, and $\epsilon^{ijk}$
is a totally anti-symmetric tensor with $\epsilon_{012}=1$.
This equation has been studied intensively 
and proved to be integrable in the sense that
dressing and splitting methods can work well
\cite{LePo}-\cite{ChLe}.

\section{Reduction to $(1+1)$-dimension}

In this section, we present various reductions
of NC ASDYM into $(1+1)$-dimensions.
In this case, two kind of translational
invariance are imposed, which is classified here
as follows:
\begin{itemize}
 \item $H_{+0}$:
       generated by $X=\del_w-\del_{\tilde{w}},~Y=\del_{\tilde{z}}$,
       which includes KdV, mKdV and NLS equations.
       The reduced NC ASDYM equation is
 \begin{eqnarray}
\label{asdym4}
&&\Phi_{\tilde{z}}^\prime 
 +[A_{\tilde{w}},\Phi_{\tilde{z}}]_\star =0,\nn
 &&
 \label{asdym5}
 \dot{\Phi}_{\tilde{z}}
 +A_{w}^\prime -A_{\tilde{w}}^\prime
 +[A_z,\Phi_{\tilde{z}}]_\star
 -[A_{w},A_{\tilde{w}}]_\star=0,\nn
 &&A_z^\prime -\dot{A}_w+[A_w,A_z]_\star =0,
\label{asdym6}  
 \end{eqnarray}
       where $(t,x)\equiv (z,w+\tilde{w})$
       and $\dot{f}:=\del f/\del t,~f^\prime:=\del f/\del
       x$.\footnote{Note that in our convention,
       $\del_x=\del_w=\del_{\tilde{w}}$. (See, Eq. (\ref{tetrad}).)}

 \item $H_{SD}$: generated by $X=\del_{\tilde{w}},~Y=\del_{\tilde{z}}$,
       which includes Boussinesq and N-wave equations,
       and topological chiral model.
       The reduced NC ASDYM equation is
\begin{eqnarray}
 \label{asdym_N}
 &&[\Phi_{\tilde{w}},\Phi_{\tilde{z}}]_\star=0,~
 A_z^\prime -\dot{A}_w+[A_w,A_z]_\star =0,\nn
&& \dot{\Phi}_{\tilde{z}}-\Phi^\prime_{\tilde{w}}
 +[A_z,\Phi_{\tilde{z}}]_\star
 -[A_{w},\Phi_{\tilde{w}}]_\star=0,
\end{eqnarray}
where $(t,x)\equiv (z,w)$.

 \item $H_{++}$: generated by $X=\del_w,~Y=\del_{\tilde{w}}$,
       which includes (affine) Toda, sine-Gordon, Liouville,
       Tzitz\'eica, and Ward's harmonic
       map and chiral equations and so on.
       The reduced NC ASDYM equation is
\begin{eqnarray}
&& \del_z{\Phi}_w+[A_z,\Phi_w]_\star=0,~ \del_{\zt}\Phi_{\wt}
  +[A_{\tilde{z}},\Phi_{\tilde{w}}]_\star=0,\nn
&&\del_z A_{\tilde{z}}-\del_{\zt} A_z
+[A_z,A_{\tilde{z}}]_\star
+[\Phi_{\tilde{w}},\Phi_w]_\star=0,
\label{++}
\end{eqnarray}
where we follow the notations
in \cite{MaWo} as well.
Only in Sec. 4.11, we choose a different choice of $X,Y$
which is defined in the relevant part.

%\item $H_{ASD}$:
\end{itemize}

\subsection{Reduction to NC KdV equation}

As we commented in Sec. 3.1,
by the identification $\del_w=\del_{\tilde{w}}$,
the reduction to NC CBS equation coincides with that \cite{Hamanaka2}
to NC KdV equation, that is, the following reduction
conditions
\begin{eqnarray}
 \label{KdV}
&&
A_{\tilde{w}}=0,~
\Phi_{\tilde{z}}=\left(\begin{array}{cc}0&0\\1&0\end{array}\right),~
A_{w}=\left(\begin{array}{cc}q&~-1\\q^\prime+q\star q
&~-q\end{array}\right),\nn
&& A_{z}=
\frac{1}{2}
\left(\begin{array}{cc}
\displaystyle q^{\prime\prime}
+2q^\prime\star q&-
2q^\prime\\
\displaystyle (1/2)q^{\prime\prime\prime}
+q^{\prime}\star q^{\prime}
+\left\{q, q^{\prime\prime}\right\}_\star 
+2q\star q^\prime \star q
&
-\displaystyle q^{\prime\prime}-2q\star q^\prime
       \end{array}\right),
\end{eqnarray}
gives rise to NC potential KdV equation
\begin{eqnarray}
 \dot{q}=\displaystyle
\frac{1}{4}q^{\prime\prime\prime}
+\frac{3}{2}\left(q\star q\right)^\prime,
\end{eqnarray}
which is derived from NC KdV equation with $u=2q^\prime$
\begin{eqnarray}
 \dot{u}=\displaystyle
\frac{1}{4}u^{\prime\prime\prime}
+\frac{3}{4}\left(u^{\prime}\star u+u\star u^{\prime}\right).
\end{eqnarray}
The present reduction belongs to $H_{+0}$ where the gauge
 group is $GL(2,\mathbb{R})$.
NC KdV equation has been studied by several authors and
proved to possess infinite conserved quantities
\cite{DiMH3} (See also \cite{Kupershmidt}.)
and exact multi-soliton solutions
\cite{EGR, Paniak}.

\vspace{3mm}

Now we comment on a gauge equivalent reduction.
First we note that
the following gauge transformation for (\ref{KdV})
leaves $\Phi_{\tilde{z}}$ as it is:
\begin{eqnarray}
 \label{gauge}
 A_\mu\rightarrow A_\mu^{(i)}=g_{(i)}^{-1}\star A_\mu \star g_{(i)}
 +g_{(i)}^{-1}\star \del_\mu g_{(i)},~~~
g_{(i)}=\left(\begin{array}{cc}1&0\\h_{(i)}&1\end{array}\right).
\end{eqnarray}
If we take $h_{(1)}=q$,\footnote{
This form is determined by the condition that
the trace elements of $A^{(1)}_w$ must vanish.}
the transformed gauge fields are calculated as 
\begin{eqnarray}
 \label{kdv2}
&&A^{(1)}_{\tilde{w}}=\left(\begin{array}{cc}0&0\\u/2&0\end{array}\right), ~
\Phi^{(1)}_{\tilde{z}}=\left(\begin{array}{cc}0&0\\1&0\end{array}\right),~
A^{(1)}_w=\left(\begin{array}{cc}0&-1\\u&0\end{array}\right),\nn
&&A^{(1)}_z=\frac{1}{4}
\left(\begin{array}{cc} u^\prime& -2u \\
 u^{\prime\prime}+2u\star u &-u^\prime
\end{array}\right),
\end{eqnarray}
with $u=2q^\prime$.
This is a gauge equivalent, direct reduction
to NC KdV equation. This coincides with the discussion
in Sec. 4.1 in \cite{Hamanaka2}
by identifying  $(\Phi^{(1)}_{\tilde{z}}
A^{(1)}_{\tilde{w}},A^{(1)}_z,A^{(1)}_w)$
with $ (\Phi_{\tilde{z}},A_{\tilde{w}},-A_z,-A_w)=(A,\tilde{U}-U,-V,-U)$
in \cite{Hamanaka2} after flipping $u\rightarrow -u$,
and with the discussion around Eq. (57) in \cite{Hamanaka_p}
by identifying $(\Phi^{(1)}_{\tilde{z}},
A^{(1)}_{\tilde{w}},A^{(1)}_z,A^{(1)}_w)$
with $(-B,-P,-H,-Q)$ in \cite{Hamanaka_p}.

We note that $U(1)$ part of gauge groups must be always taken into
account even when all gauge fields are traceless in some gauge
as in (\ref{kdv2}) because gauge transformations could always give rise
to trace parts. For example, explicit calculation of
the gauge transformation (\ref{gauge}) is
\begin{eqnarray*}
 \left(\begin{array}{cc}1&0\\h&1\end{array}\right)^{-1}\star
 \left(\begin{array}{cc}a&b\\c&d\end{array}\right)\star
 \left(\begin{array}{cc}1&0\\h&1\end{array}\right)=
 \left(\begin{array}{cc}a+b\star h &0\\c+d\star h-h\star a -h\star
       b\star h &d-h\star b\end{array}\right).
\end{eqnarray*}
The trace increases by $[b,h]_\star$ after the transformation.
The $U(1)$ part actually plays important roles
in NC gauge theories and gives rise to new physical objects
\cite{NC, NeSc, YM}.

\subsection{Reduction to NC mKdV equation}

Here let us consider a symmetry reduction $H_{+0}$:
$X=\del_w-\del_{\tilde{w}},~Y=\del_{\tilde{z}}$
where the gauge group is $GL(2,\mathbb{R})$.

Now let us take further reduction condition on gauge fields:
\begin{eqnarray}
 \label{mkdv}
 \Phi_{\tilde{z}}=
  \left(\begin{array}{cc}0&0\\1&0\end{array}\right),~
 A_{\tilde{w}}=
  \left(\begin{array}{cc}0&0\\a&0\end{array}\right),~
 A_z=
  \left(\begin{array}{cc}c&b\\0&d\end{array}\right),~
 A_{w}=
  \left(\begin{array}{cc}v&-1\\0&-v\end{array}\right).
\end{eqnarray}
The first equation of Eq. (\ref{asdym4}) is automatically satisfied.
Eq. (\ref{asdym5}) becomes
\begin{eqnarray}
  \left(\begin{array}{cc}
        v^\prime +a+b&0\\
        -a^\prime+d-c+\left\{a, v\right\}_\star&~-v^\prime-a-b
        \end{array}\right)=0,
\end{eqnarray}
hence, $b=-v^\prime-a,~c-d=-a^\prime+a\star v+v\star a$.
The second equation of Eq. (\ref{asdym6}) becomes
\begin{eqnarray}
  \left(\begin{array}{cc}
        c^\prime-\dot{v}+[v,c]_\star & b^\prime+c-d+\left\{b,v\right\}_\star\\
        0&d^\prime+\dot{v}-[v,d]_\star
        \end{array}\right)=0.
  \label{2}
\end{eqnarray}
{}From the 1-2 component of Eq. (\ref{2}), we get
\begin{eqnarray}
 a=-\frac{1}{2}v^\prime -\frac{1}{2}v\star v,
\end{eqnarray}
and hence
\begin{eqnarray}
 b=-\frac{1}{2}v^\prime +\frac{1}{2}v\star v,~
 c-d=\frac{1}{2}v^{\prime\prime}-v\star v\star v.
\end{eqnarray}
{}From the trace of Eq. (\ref{2}),
we get $c+d=-(1/2)[v,v^\prime]_\star$. Therefore
\begin{eqnarray}
 c=\frac{1}{4}v^{\prime\prime}-\frac{1}{2}v\star v\star v
  -\frac{1}{4}[v,v^\prime]_\star,~
 d=-\frac{1}{4}v^{\prime\prime}+\frac{1}{2}v\star v \star v
 -\frac{1}{4}[v,v^\prime]_\star.
\end{eqnarray}
Finally we get NC mKdV eq from the diagonal elements of Eq. (\ref{2}): 
\begin{eqnarray}
 \dot{v}=\frac{1}{4}v^{\prime\prime\prime}
  -\frac{3}{4}\left\{v\star v, v^{\prime}\right\}_\star,
\end{eqnarray}
which is connected with NC KdV equation
via NC Miura map: $u=v^\prime -v^2$ \cite{DiMH3}:
\begin{eqnarray}
 \dot{u}=\frac{1}{4}u^{\prime\prime\prime}
  +\frac{3}{4}(u\star u^{\prime}
  +u^{\prime}\star u).
\end{eqnarray}

\vspace{3mm}

We comment that this reduction is gauge equivalent to
that of NC KdV equation
under the gauge transformation
(\ref{gauge}) with\footnote{
The explicit form of $h_{(2)}$ and the relationship between $q$ and $v$
(just the NC Miura map) are determined by the condition
that $A^{(2)}_w$ must coincide with $A_w$ in (\ref{mkdv}).}
\begin{eqnarray}
h_{(2)}=-\frac{1}{2}\left(v+\del_x^{-1}(v\star v)\right),~~~
(u=)~2q^\prime=v^\prime-v^2.
\end{eqnarray}
The transformed gauge fields are calculated as 
\begin{eqnarray}
&&\Phi^{(2)}_{\tilde{z}}=\left(\begin{array}{cc}0&0\\1&0\end{array}\right),~
A^{(2)}_{\tilde{w}}=\frac{1}{2}\left(\begin{array}{cc}0&0\\
-v^\prime -v\star v &0\end{array}\right),~
A^{(2)}_w=\frac{1}{4}
\left(\begin{array}{cc} v& -1 \\
 0&-v
\end{array}\right),\nn
&&A^{(2)}_z=\frac{1}{4}
\left(\begin{array}{cc}v^{\prime\prime}
 -2v\star v\star v-[v,v^\prime]_\star&
 -2v^{\prime}+2v\star v\\
 0&-v^{\prime\prime}+2v\star v\star v-[v,v^\prime]_\star\end{array}\right).
\end{eqnarray}
This is just the present reduction to NC mKdV
which has infinite conserved quantities \cite{DiMH3}.

\subsection{Reduction to NC NLS equation}

As we commented in Sec. 3.2,
by the identification $\del_w=\del_{\tilde{w}}$,
the reduction to NC Zakharov system coincides with
that \cite{Legare, Hamanaka2} to NC NLS equation,
that is, the following reduction
conditions
\begin{eqnarray}
 \label{nls}
\Phi_{\tilde{z}}=
\frac{i}{2}\left(\begin{array}{cc}1&0\\0&-1\end{array}\right),~
A_{\tilde{w}}=0,~
A_{w}=\left(\begin{array}{cc}0&\psi\\
\varepsilon\bar{\psi}&0\end{array}\right),~
A_{z}=i\varepsilon\left(\begin{array}{cc}
-\psi\star\bar{\psi}
&-\varepsilon\psi^\prime\\
\bar{\psi}^\prime&
\bar{\psi}\star\psi
\end{array}\right),
\end{eqnarray}
gives rise to NC NLS equation
\begin{eqnarray}
 \label{nls}
i\dot{\psi}=\psi^{\prime\prime}-2\varepsilon\psi \star \bar{\psi} \star \psi
\end{eqnarray}
where $\varepsilon=\pm 1$.
The NC NLS equation (\ref{nls})
possesses infinite conserved quantities \cite{DiMH2}.

The KdV and NLS equations and the hierarchies 
possess the twistor descriptions \cite{MaSp, MaSp2}.
NC extension of them are very interesting and
the details will be reported later.
 
\subsection{Reduction to NC Boussinesq equation}

Here we consider a symmetry reduction $H_{SD}$:
$X=\del_{\tilde{w}},~Y=\del_{\tilde{z}}$
where the gauge group is $GL(3,\mathbb{R})$.
%Under the gauge $\Phi_w=\Phi_z=0$,
The reduced ASDYM is the same as Eq. (\ref{asdym_N}):
Now let us take further reduction condition on gauge fields:
\begin{eqnarray*}
 \Phi_{\tilde{z}}=
  \left(\begin{array}{ccc}0&0&0\\0&0&0\\1&0&0\end{array}\right),~
 \Phi_{\tilde{w}}=
  \left(\begin{array}{ccc}0&0&0\\1&0&0\\0&1&0\end{array}\right),~
 A_z=
  \left(\begin{array}{ccc}a&0&-1\\d&b&0\\f&e&c\end{array}\right),~
 A_{w}=
  \left(\begin{array}{ccc}0&-1&0\\0&0&-1\\v&u&0\end{array}\right),
\end{eqnarray*}
where all matrices are traceless.
Eq. (\ref{asdym_N}) gives rise to the following equations:
\begin{eqnarray}
&&a-b+u=0, ~b-c=0,~a^\prime-d+v=0,~b^\prime-e+d=0,~
 c^\prime-v+e=0,~
 d^\prime-f=0,\nn&&
 f^\prime-\dot{v}+v\star a
+u\star d -c\star v=0, ~
e^\prime-\dot{u}+u\star b+f-c\star u=0.
\end{eqnarray}
Together with the traceless condition $a+b+c=0$, we get
\begin{eqnarray}
a=-\frac{2}{3}u,~b=c=\frac{1}{3}u,~d=-\frac{2}{3}u^\prime+v,~
e=-\frac{1}{3}u^\prime+v,~
f=-\frac{2}{3}u^{\prime\prime}+v^\prime.
\end{eqnarray}
The remaining equations are
\begin{eqnarray}
\dot{v}=-\frac{2}{3}u^{\prime\prime\prime}+v^{\prime\prime}
 -\frac{2}{3}u\star u^\prime+\frac{2}{3}[u,v]_\star,~~
 \dot{u}=-u^{\prime\prime}+2v^\prime.
\end{eqnarray}
Eliminating $v$, we get NC Boussinesq equation \cite{Toda}:
\begin{eqnarray}
\ddot{u}+\frac{1}{3}u^{\prime\prime\prime\prime}+\frac{2}{3}
 (u\star u)^{\prime\prime}-\frac{2}{3}([u,\del_x^{-1}\dot{u}]_\star)^\prime=0.
\end{eqnarray}
This coincides with that derived from NC Gelfand-Dickey hierarchy
\cite{HaTo3} and therefore possesses infinite
conserved quantities \cite{Hamanaka}
and exact multi-soliton solutions \cite{EGR}.

The potential form (NC potential Boussinesq equation)
is obtained by setting $u=q^\prime$:
\begin{eqnarray}
\ddot{q}+\frac{1}{3}q^{\prime\prime\prime\prime}+\frac{2}{3}
\left\{q^\prime, q^{\prime\prime}\right\}_\star
+\frac{2}{3}[\dot{q},q^\prime]_\star=0.
\end{eqnarray}

In similar way, we can embed NC version of
Drinfeld-Sokolov reductions \cite{DrSo} into a framework of
NC ASDYM equation or NC twistor theory
as in commutative case \cite{MaWo, MaSi, BaDe}.
The detailed discussion will be reported later.

\subsection{Reduction to NC N-wave equation}

Here we consider a symmetry reduction $H_{SD}$:
$X=\del_{\tilde{w}},~Y=\del_{\tilde{z}}$
where the gauge group is $GL(N,\mathbb{R})$.
If we take
\begin{eqnarray}
 \Phi_{\tilde{w}}=\diag(a_1,\cdots, a_N),~
 \Phi_{\tilde{z}}=\diag(b_1,\cdots, b_N),~
 (A_w)_{ij}=\omega_{ij},~(A_z)_{ij}=\lambda_{ij}\omega_{ij},
\end{eqnarray}
where $a_i, b_i,(i=1,\cdots N)$ are constants and
$\dis \lambda_{ij}:=(b_i-b_j)/(a_i-a_j)=\lambda_{ji},
~(i\neq j),~\lambda_{ii}=0$,
and $\omega_{ij}=-\omega_{ji}$,
we get NC $N$-wave equation \cite{Hamanaka2}:
\begin{eqnarray}
 \dot{\omega}_{ij}=
 \lambda_{ij}\omega_{ij}^\prime
 -\sum_{k=1}^{N}(\lambda_{ik}-\lambda_{kj})\omega_{ik}\star \omega_{kj}.
\end{eqnarray}
This reduction has the symmetry of reflection
$\rho: w\mapsto w, z\mapsto z, \wt\mapsto -\wt, \zt \mapsto -\zt$.
For the commutative reduction, see e.g. \cite{MaWo,AbCh}.

\subsection{Reduction to NC topological chiral model}

Here we consider a symmetry reduction $H_{SD}$:
$X=\del_{\tilde{w}},~Y=\del_{\tilde{z}}$
where the gauge group is $GL(N,\mathbb{C})$.
First we take a gauge in which $A_w=A_z=0$.
Then the linear systems for NC ASDYM equation
becomes $L=\del_x+\Phi_{\zt},~M=\del_t +\Phi_{\wt}$
at $\lambda=-1$, which implies that
\begin{eqnarray}
 \Phi_{\wt}=g^{-1}\star \del_t g,~~~
 \Phi_{\zt}=g^{-1}\star \del_x g,
\end{eqnarray}
where $g(t,x)$ is a nonsingular
matrix-valued function of $x$ and $t$.
Then NC ASDYM equation is equivalent to
\begin{eqnarray}
 \del_x(g^{-1}\star \del_t g)
 -\del_t(g^{-1}\star \del_x g)=0,
\end{eqnarray}
which is a NC version of topological chiral model \cite{MaWo}.
This equation can be expressed as $d(g^{-1}\star dg)=0$.
%and hence does not involve a metric.

\subsection{Reduction to NC (affine) Toda field equations}

Here we consider a symmetry reduction $H_{++}$:
$X=\del_w=0,~Y=\del_{\tilde{w}}=0$
where the gauge group is $GL(N,\mathbb{C})$.
If we take
\begin{eqnarray}
 A_z= \left(\begin{array}{cccc}a_1&&&O\\&a_2&&\\&&\ddots&\\O&&&a_N
              \end{array}\right),~
% \diag(a_1,\cdots, a_N),~
 A_{\tilde{z}}=\left(\begin{array}{cccc}-\tilde{a}_1&&&O\\&-\tilde{a}_2&&\\
                      &&\ddots&\\O&&&-\tilde{a}_N
              \end{array}\right),~\nn
%\diag(-\tilde{a}_1,\cdots, -\tilde{a}_N),~\nn
 \Phi_w=\left(\begin{array}{ccccc}0&\phi_1&&&O\\&0&\phi_2&&\\
                 &&0&\ddots&\\&O&&\ddots&\phi_{N-1}\\\epsilon\phi_N&&&&0
              \end{array}\right),~
%(\delta_{i,j-1}+\epsilon \delta_{i,j+N-1})\phi_i,~
 \Phi_{\tilde{w}}=\left(\begin{array}{ccccc}0&&&&\epsilon\tilde{\phi}_N\\
                \tilde{\phi}_1&0&&O&\\
                 &\tilde{\phi}_2&0&&\\&&\ddots&\ddots&\\O&&&\tilde{\phi}_{N-1}&0
              \end{array}\right),~
%(\delta_{i,j+1}+\epsilon \delta_{i,j-N+1})\tilde{\phi}_i,
\end{eqnarray}
where $\epsilon$ is a real constant which values 0 or 1,
we get NC Toda equation for $\epsilon=0$ \cite{Lee}:
%(the suffix $i$ runs $i=1,2,\cdots,N-1$ for NC Toda equation
%and $i=1,2,\cdots,N$ for affine Toda equation):
\begin{eqnarray}
&& \del_z{\phi}_i=\phi_i\star a_{i+1}-a_i\star \phi_i, ~~~
\del_{\zt}\tilde{\phi}_i=\tilde{a}_{i+1}\star \phit_i
-\phit_i\star \tilde{a}_{i}, ~~~(i=1,2,\cdots,N-1)\nn
&&\del_z{\tilde{a}}_1+\del_{\zt}a_1+[a_1,\at_1]_\star
+\phi_1\star \tilde{\phi}_1=0,\nn
&&\del_z{\tilde{a}}_i+\del_{\zt}a_i+[a_i,\at_i]_\star
+\phi_i\star \tilde{\phi}_i
-\phi_{i-1}\star \tilde{\phi}_{i-1}=0,~~~(i=2,3,\cdots,N-1)\nn
&&\del_z{\tilde{a}}_N+\del_{\zt}a_N+[a_{N},\at_N]_\star
-\phi_{N-1}\star \tilde{\phi}_{N-1}=0,
\end{eqnarray}
and NC affine Toda equation for $\epsilon=1$ \cite{Lee}:
\begin{eqnarray}
&& \del_z{\phi}_i=\phi_i\star a_{i+1}-a_i\star \phi_i, ~~~
\del_{\zt}\tilde{\phi}_i=\tilde{a}_{i+1}\star \phit_i
-\phit_i\star \tilde{a}_{i}, ~~~(i=1,2,\cdots,N-1)\nn
&& \del_z{\phi}_N=\phi_N\star a_1-a_N\star \phi_N, ~~~
\del_{\zt}\tilde{\phi}_N=\tilde{a}_1\star \phit_N
-\phit_N\star \tilde{a}_N,\nn
&&\del_z{\tilde{a}}_1+\del_{\zt}a_1+[a_1,\at_1]_\star
+\phi_1\star \tilde{\phi}_1-\phi_{N}\star \tilde{\phi}_{N}=0,\nn
&&\del_z{\tilde{a}}_i+\del_{\zt}a_i+[a_i,\at_i]_\star
+\phi_i\star \tilde{\phi}_i
-\phi_{i-1}\star \tilde{\phi}_{i-1}=0.~~~(i=2,3,\cdots,N).
\end{eqnarray}
The NC (affine) Toda field equations have
exact soliton solutions \cite{Lee}.
We note that the NC (A)SD Chern-Simons equation
coupled to an adjoint matter just coincides with
Eq. (\ref{++}).
For $N=2$, the NC Toda equation and the NC affine Toda equation
include NC Liouville equation and
NC sinh-Gordon equation, respectively \cite{Lee}.
In the commutative limit, these equations reduce to
ordinary form of (affine) Toda equations:
\begin{eqnarray}
 \del_z\del_{\zt}u_i+\sum_{j}K_{ij}e^{u_j}=0,
\end{eqnarray}
where $u_i=\log (\phi_i\phit_i)$ and
the matrix $K_{ij}$ is an (extended) Cartan matrix
associated to $SU(N)$.
(The explicit definition is seen in e.g. \cite{MaWo}.)

This reduction possesses an additional symmetry an in commutative
case \cite{MaWo}. Let us take the rotational transformation
$w\rightarrow w'=e^{i\theta} w$. Then the one-form $\Phi_w dw$
is transformed to $\Phi'_wdw'=G_\theta^{-1} \Phi_w G_\theta dw'$
where $G_\theta=\diag (1,\alpha^{-1},
\cdots, \alpha^{-(N-1)})$ and $\alpha=e^{i\theta}$.
$\Phi_w'$ is calculated as 
\begin{eqnarray}
 \Phi_w'=\alpha^{-1}\left(\begin{array}{ccccc}0&\phi_1&&&O\\&0&\phi_2&&\\
                 &&0&\ddots&\\&O&&\ddots&\phi_{N-1}\\
              \alpha^N \epsilon\phi_N&&&&0
              \end{array}\right).
\end{eqnarray}
Hence the one-form has the rotational symmetry
generated by $Z=iw\del_w-i\wt\del_{\wt}$ for $\epsilon=0$, 
and the discrete rotational symmetry where $\theta$ is integer
multiples of $(2\pi/N)$ for $\epsilon=1$.
For the commutative reduction, see e.g. \cite{MaWo, AbCh}.

\subsection{Reduction to NC sine-Gordon equations}

In this subsection, we review the reduction of
NC Yang's equation to NC sine-Gordon equation
where the gauge group is $GL(2,\mathbb{C})$.
For the commutative discussion, see \cite{GWW}. 
In such reductions, exponential functions are involved and 
NC extension of them has several possibilities.
Naive definition of NC exponential functions is as follows
\begin{eqnarray}
 e_\star^{\phi}:=\sum_{n=0}^{\infty}\frac{1}{n!}
  \underbrace{\phi\star
  \cdots \star \phi}_{n \mbox{\scr{ times}}}.
\end{eqnarray}

Let us start with the NC Yang's equation (\ref{yang}) 
and take the following ansatz:
\begin{eqnarray}
 J=e_\star^{i\sigma_1 \wt}\star g(z,\zt)\star e_\star^{i\sigma_1 w}.
\end{eqnarray}
Then NC Yang's equation (\ref{yang}) reduces to
\begin{eqnarray}
\label{yang2}
 \del_z(g^{-1} \star \del_{\zt} g)-
 \left[\sigma_1, ~g^{-1}\star\sigma_1 g\right]_\star=0.
\end{eqnarray}
This equation does not depend on $w$ and $\wt$
and the present reduction is considered to belong to $H_{++}$.
Pauli matrices are defined as usual:
\begin{eqnarray}
%  \sigma_0=
%  \left(\begin{array}{cc}1&0\\0&1\end{array}\right),~
 \sigma_1=
  \left(\begin{array}{cc}0&1\\1&0\end{array}\right),~
 \sigma_2=
  \left(\begin{array}{cc}0&-i\\i&0\end{array}\right),~
 \sigma_3=
  \left(\begin{array}{cc}1&0\\0&-1\end{array}\right).
\end{eqnarray}
{}From this equation, we can get two kind of
NC sine-Gordon equations:
\begin{eqnarray}
 \del_{z}\left(e_\star^{-iu/2}\star \del_{\zt} e_\star^{iu/2}
 \right)
 &=&2i\sin_\star u\nn
 \del_{z}\left(e_\star^{iu/2}\star \del_{\zt} e_\star^{-iu/2}\right)
 &=&-2i\sin_\star u
\end{eqnarray}
for $g=\exp_\star\left\{(i/2)\sigma_3 u \right\}$ \cite{GMPT}, and
\begin{eqnarray}
 \label{sG2}
 \del_{z}\left(e_\star^{-iu/2}\star \del_{\zt} e_\star^{iu/2}
 \right)+
 \del_{z}\left(e_\star^{-iu/2}\star
 V\star e_\star^{iu/2}\right)&=&
 2i\sin_\star u\nn
 \del_{z}\left(e_\star^{iu/2}\star \del_{\zt} e_\star^{-iu/2}
\right)+\del_{z}\left(e_\star^{iu/2}\star
V\star  e_\star^{-iu/2}\right)
 &=&
 -2i\sin_\star u,
\end{eqnarray}
for $g=\exp_\star\left\{(i/2)  v\right\}\star
\exp_\star\left\{(i/2) \sigma_3 u\right\}$ \cite{LMPPT},
where
\begin{eqnarray}
 \sin_\star u:=\frac{e_\star^{iu}-e_\star^{-iu}}{2i},~
 V:=  e_\star^{-iv/2}\star \del_{\zt} e_\star^{iv/2}.
\end{eqnarray}
Both of them possess infinite conserved quantities
\cite{GrPe, GMPT, LMPPT}
and the latter preserves,
due to existence of two kind of fields,
an integrable property of
factorized S-matrix \cite{LMPPT}.

%\begin{eqnarray}
%&& \dot{\phi}_1=-a_1\star \phi_1+\phi_1\star a_{2}, ~~~
%\tilde{\phi}^\prime_1=\tilde{a}_1\star \phit_1
%-\phit_1\star \tilde{a}_{2}, \nn
%&& \dot{\phi}_2=-a_2\star \phi_2+\phi_2\star a_{1}, ~~~
%\tilde{\phi}^\prime_2=\tilde{a}_2\star \phit_2
%-\phit_2\star \tilde{a}_{1}, \nn
%&&\dot{\tilde{a}}_1+a_1^\prime+[a_1,\at_1]_\star
%+\phi_1\star \tilde{\phi}_1-\phi_{2}\star \tilde{\phi}_{2}=0,\nn
%&&\dot{\tilde{a}}_2+a_2^\prime+[a_2,\at_2]_\star
%+\phi_2\star \tilde{\phi}_2
%-\phi_{1}\star \tilde{\phi}_{1}=0.
%\end{eqnarray}
In the commutative limit, both of them reduce to
sine-Gordon equation:
\begin{eqnarray}
 \del_z\del_{\zt}u=4\sin u.
\end{eqnarray}
For (\ref{sG2}), there is an additional
decoupled equation $\del_z\del_{\zt}v=0$.

\subsection{Reduction to NC Liouville equation}

In this subsection, we review the reduction of
NC Yang's equation to NC Liouville equation by
Cabrera-Carnero \cite{Cabrera-Carnero},
where the gauge group is $GL(2,\mathbb{R})$.
For the commutative discussion, see \cite{GWW}.

Let us start with the NC Yang's equation (\ref{yang}) 
and take the following ansatz:
\begin{eqnarray}
 J=e_\star^{\sigma_- \wt} \star g(z,\zt)\star e_\star^{-\sigma_+ w}.
\end{eqnarray}
Then NC Yang's equation (\ref{yang}) reduces to
\begin{eqnarray}
\label{yang2}
 \del_z(g^{-1} \star \del_{\zt} g)-
 \left[\sigma_-, ~g^{-1}\star\sigma_+ g\right]_\star=0,
\end{eqnarray}
where $\sigma_\pm:=(1/2)(\sigma_1\pm i\sigma_2)$ and 
the present reduction is considered to belong to $H_{++}$
for the same reason as the case for NC sine-Gordon equation.
{}From this equation, we can get a NC Liouville equation:
\begin{eqnarray}
 \del_{z}\left(e_\star^{-\phi_+}\star \del_{\zt}e_\star^{\phi_+}
 \right)
 &=&e_\star^{-\phi_-}\star e_\star^{\phi_+}\nn
 \del_{z}\left(e_\star^{-\phi_-}\star \del_{\zt}e_\star^{\phi_-}\right)
 &=&- e_\star^{-\phi_-}\star e_\star^{\phi_+},
\end{eqnarray}
for $g=\diag(\exp_\star\left\{\phi_+\right\},
\exp_\star\left\{\phi_-\right\})$
where $\phi_+=v+u, ~\phi_-=v-u$ \cite{Cabrera-Carnero}.
This equation also has soliton solutions \cite{Cabrera-Carnero2}. 

%\begin{eqnarray}
%&& \dot{\phi}_1=-a_1\star \phi_1+\phi_1\star a_{2}, ~~~
%\tilde{\phi}^\prime_1=\tilde{a}_1\star \phit_1
%-\phit_1\star \tilde{a}_{2}, \nn
%&&\dot{\tilde{a}}_1+a_1^\prime+[a_1,\at_1]_\star
%+\phi_1\star \tilde{\phi}_1=0,~~~
%\dot{\tilde{a}}_2+a_2^\prime+[a_2,\at_2]_\star
%-\phi_{1}\star \tilde{\phi}_{1}=0.
%\end{eqnarray}
In the commutative limit, this reduce to
Liouville equation:
\begin{eqnarray}
 \del_z\del_{\zt}u=e^{2u},
\end{eqnarray}
together with a decoupled equation $\del_z\del_{\zt} v=0$.

\subsection{Reduction to NC Tzitz\'eica equation}

Here let us discuss the reduction of
NC Yang's equation to a NC Tzitz\'eica equation
where the gauge group is $GL(3,\mathbb{R})$, which is new.

Let us start with the NC Yang's equation (\ref{yang}) 
and take the following ansatz:
\begin{eqnarray}
 J=e_\star^{-E_- \wt}\star g(z,\zt)\star e_\star^{E_+ w},
\end{eqnarray}
where
\begin{eqnarray}
E_-=\left(\begin{array}{ccc}0&0&1\\1&0&0\\0&1&0\end{array}\right),~~~
E_+=\left(\begin{array}{ccc}0&1&0\\0&0&1\\1&0&0\end{array}\right).
\end{eqnarray}
Then NC Yang's equation (\ref{yang}) reduces to
\begin{eqnarray}
\label{yang2}
 \del_z(g^{-1} \star \del_{\zt} g)-
 \left[E_-, ~g^{-1}\star E_+ g\right]_\star=0.
\end{eqnarray}
As in Secs 4.8 and 4.9,
the present reduction belongs to $H_{++}$,
and leads to a NC version of Tzitz\'eica equation
by taking $g=e_\star^{\rho}\star \diag(e_\star^\omega,
e_\star^{-\omega}, 1)$: 
\begin{eqnarray}
 \label{Tz}
&& \del_{z}\left(e_\star^{-\omega}\star \del_{\zt} e_\star^{\omega}
 \right)+
 \del_{z}\left(e_\star^{-\omega}\star
 V\star e_\star^{\omega}\right)=
 e_\star^\omega -e_\star^{-2\omega}\nn
&& \del_{z}\left(e_\star^{\omega}\star \del_{\zt} e_\star^{-\omega}
\right)+\del_{z}\left(e_\star^{\omega}\star
V\star e_\star^{-\omega}\right)
 =e_\star^{-2\omega}-e_\star^{\omega},\nn
&&\del_z V=0,
\end{eqnarray}
where $V:=e_\star^{-\rho}\star \del_{\zt} e_\star^{\rho}$.
The present reduction is a special case of reduction
to NC affine Toda equations \cite{Cabrera-Carnero}
and hence NC Tzitz\'eica equation
possesses the same kind of good properties
as NC affine Toda equation \cite{Cabrera-Carnero, Lee}.

In the commutative limit, these equations reduce to
Tzitz\'eica equation:
\begin{eqnarray}
 \del_z\del_{\zt}\omega=e^\omega-e^{-2\omega}.
\end{eqnarray}
together with a decoupled equation $\del_z\del_{\zt}\rho=0$.
The present result is different from that by Dunajski \cite{Dunajski}.

\subsection{Reduction to NC Ward's chiral and harmonic map equations}

Here we discuss two real reductions of NC Yang's equation (\ref{yang})
by $H_{++}:$ $X=\del_z-\del_{\wt}, Y=\del_w+\del_{\zt}$. 
By defining $u=z+\wt, v=w-\zt$, we get the following equation:
\begin{eqnarray}
 \label{red_yang}
 \del_u(J^{-1} \star \del_{v} J)+\del_v (J^{-1}\star \del_{u} J)=0.
\end{eqnarray}
As in commutative case, we obtain two kind of
reductions by imposing two different reality conditions
for ultrahyperbolic signature.
\begin{itemize}
 \item NC Ward's harmonic map equation 

       By putting $\bar{w}=z, \bar{\zt}=-\zt$, so that $X=\bar{Y}$
       and $u=\bar{v}$, the reduced equation is NC Ward's harmonic map
       equation: $J: \mathbb{C}\rightarrow G$:
\begin{eqnarray}
 \del_u(J^{-1} \star \del_{\bar{u}} J)
  +\del_{\bar{u}} (J^{-1}\star \del_{u} J)=0,
\end{eqnarray}
with the metric $ds^2=2(dzd\zt+d\bar{z}d\bar{\zt})$.
This is a simple NC version of
Ward's harmonic map equation \cite{Ward}.

\item NC Ward's chiral equation in $(1+1)$-dimension

      By taking $w,z,\wt,\zt$ to be real, that is $u,v$ to be also real,
      Eq. (\ref{red_yang})
      is just the NC version of Ward's chiral equation \cite{Ward}.
\end{itemize}
For Euclidean signature, there is another choice of reality condition
which leads to a NC version of harmonic map equation \cite{MaWo}.
These equations have not yet been examined in detail.
(See also \cite{LPS2, ChLe}.)

\section{Conclusion and Discussion}

In the present paper, we proved that
various NC integrable equations in both
$(2+1)$ and $(1+1)$ dimensions
are actually derived from NC ASDYM equations
in the ultrahyperbolic signature
by reduction. Existence of these reductions guarantees that
the lower-dimensional integrable equations actually have 
both corresponding physical situations, such as, reduced D0-D4
D-brane systems, and twistor descriptions of them.
We can expect that analysis of exact NC soliton solutions
could be applied to that of  D-brane dynamics,
and NC twistor theory would explain
geometrical origin of NC integrable equations

There are mainly two steps to go further.
One is to develop twistor descriptions of
the present results. For example, twistor descriptions of
KdV and NLS equations are given by Mason and Sparling \cite{MaSp, MaSp2}.
In their work, the relationship between Penrose-Ward 
transformation and inverse scattering method
has been revealed.
Inverse scattering method is one of the most traditional way
to discuss their integrability in terms of action-angle
variables. NC Penrose-Ward transformation for NC ASDYM equation
has been already developed
by Kapustin, Kuznetsov and Orlov \cite{KKO}, and
Brain and Hannabuss \cite{Hannabuss, Brain, BrHa}.
Hence NC extension of them could be possible
and worth studying
and will be reported later somewhere.
Generalization of such twistor frameworks to $(2+1)$-dimension
is developed by Strachan \cite{Strachan2} and
the NC extension is also interesting. In this case,
noncommutativity could be introduced into
space directions only and integrability could be
defined as usual. In this sense, the twistor descriptions
of them are expected to be easy to treat.
Twistor interpretation of NC Drinfeld-Sokolov reductions 
and NC B\"acklund transformation presented in Sec. 2.4
should be developed to reveal a hidden symmetry of
NC integrable hierarchies.
It is also interesting to study a connection
between the twistor description of the NC hierarchy and
non-associative structure behind NC integrable systems
developed by Dimakis and M\"uller-Hoissen \cite{DiMH8}.

Another direction is to reveal the corresponding
physical situations in N=2 string theory and
to apply exact analysis of NC integrable equations
to them. The BPS equations in some D-brane configurations
would just correspond to NC integrable equations
and the soliton solutions correspond to
lower-dimensional D-branes.
In these situations, we can expect that
similar applications to D-brane dynamics
would be possible and successful.
NC dressing and splitting methods would be also
applicable to NC KdV equation and so on.
NC solitons are sometimes so easy to treat
that we could analyze energy densities of them
and fluctuations around them and so on.
This would be a hint to reveal
the corresponding D-brane configurations which might
be new BPS states such as \cite{EINOOST}.

\subsection*{Acknowledgments}

The author would like to give special thanks to L.~Mason 
for a lot of fruitful discussions and helpful comments.
He is also grateful to F.~M\"uller-Hoissen
for valuable comments via e-mail correspondences.
Thanks are also due to
C.~Athorne, C.~S.~Chu, E.~Corrigan, M.~Dunajski,
K.~Hashimoto, N.~Manton,
I.~Strachan, R.~Szabo and C.~A.~S.~Young
for hospitality and discussion
during stays at universities of York,
Heriot-Watt, Glasgow, Durham and Cambridge.
This work was supported
by the Yamada Science Foundation
for the promotion of the natural science.

%\begin{appendix}
%\section{Miscellaneous Formulas}
%\end{appendix}

\end{document}